\documentclass{Interspeech2024}

\interspeechcameraready

\title{Diversifying and Expanding Frequency-Adaptive Convolution Kernels \\ for Sound Event Detection}

\name[affiliation={1}]{Hyeonuk}{Nam}
\name[affiliation={2}]{Seong-Hu}{Kim}
\name[affiliation={1}]{Deokki}{Min}
\name[affiliation={3}]{Junhyeok}{Lee}
\name[affiliation={1}]{Yong-Hwa}{Park}

\address{
  $^1$Korea Advanced Institute of Science and Technology, South Korea,\\
  $^2$Samsung Research, South Korea, 
  $^3$Supertone Inc, South Korea}
\email{frednam@kaist.ac.kr, seonghu.kim@samsung.com, minducky@kaist.ac.kr, jun.hyeok@supertone.ai, yhpark@kaist.ac.kr}


\begin{document}

\maketitle

\begin{abstract}
Frequency dynamic convolution (FDY conv) has shown the state-of-the-art performance in sound event detection (SED) using frequency-adaptive kernels obtained by frequency-varying combination of basis kernels. However, FDY conv lacks an explicit mean to diversify frequency-adaptive kernels, potentially limiting the performance. In addition, size of basis kernels is limited while time-frequency patterns span larger spectro-temporal range. Therefore, we propose \textit{dilated frequency dynamic convolution (DFD conv)} which diversifies and expands frequency-adaptive kernels by introducing different dilation sizes to basis kernels. Experiments showed advantages of varying dilation sizes along frequency dimension, and analysis on attention weight variance proved dilated basis kernels are effectively diversified. By adapting class-wise median filter with intersection-based F1 score, proposed DFD-CRNN outperforms FDY-CRNN by 3.12\% in terms of polyphonic sound detection score (PSDS).
\end{abstract}

\section{Introduction}
\label{sec:intro}
Sound event detection (SED) which aims to classify the target sound events with corresponding time location within a given audio clip is one of the most active research themes to reproduce auditory intelligence. By reproducing auditory intelligence, we could automate the functions of human auditory system for potential applications such as robotics, automation, surveillance, media information retrieval, etc. \cite{CASSE, DCASEtask4, crnn, sedmetrics, mytechreport}. While most of works on SED have been directly applying methods from other fields such as image recognition and speech recognition, there exist recent works those consider and apply characteristics of acoustics and sound events into SED methods \cite{FDY, filtaug, freqatt}.

Frequency dynamic convolution (FDY conv) is one representative example, motivated by the fact that sound event information is shift-variant on frequency dimension. FDY conv applies convolution kernel that varies along frequency dimension \cite{FDY} by using frequency-adaptive attention weights to perform weighted sum on basis kernels to obtain frequency-adaptive convolution kernels. SED model with FDY conv has achieved state-of-the-art performance on domestic environment sound event detection (DESED) real validation dataset \cite{DCASEtask4, crnn, FDY, dcase2023_1st, dcase2023_2nd}.

However, there still is a room for improvements on FDY conv. The size of basis kernels is fixed as 3 by 3, while time-frequency patterns by sound events span larger spectro-temporal range. While convolutional recurrent neural network (CRNN) structure considers long temporal range using gated recurrent unit (GRU) module, it neglects the importance of recognizing broad spectral range. In addition, it is difficult to guarantee the diversity of multiple basis kernels, as all basis kernels are structurally the same. Thus diversity of frequency-adaptive kernels over frequency dimension might be limited. To remedy aforementioned points, we aim to improve FDY conv by expanding the spectral receptive field and diversifying frequency-adaptive kernels by assigning different roles to the basis kernels.
The main contributions of this work are as follows:
\begin{enumerate}
    \itemsep0.1em
    \item{Proposed \textit{dilated frequency dynamic convolution (DFD conv)} applies varying dilation sizes to FDY conv basis kernels to diversify frequency-adaptive kernels and expand their spectral receptive field.}
    \item{Experimental results proved the effectiveness of dilating basis kernels along frequency dimension, and advantage of applying varying dilation sizes to different basis kernels.}
    \item{By comparing attention weight vector variance of FDY conv and DFD conv, we showed that frequency-adaptive attention weights are more efficiently diversified in DFD conv.}
    \item{Proposed DFD-CRNN improved baseline FDY-CRNN by 2.43\%  and adapting class-wise median filter outperformed previous state-of-the-art FDY-CRNN by 3.12\% by polyphonic sound detection score (PSDS) \cite{PSDS}.}
\end{enumerate}



\section{Methods}
\label{sec:methods}
\begin{figure}[t]
\centerline{\includegraphics[width=7.5cm]{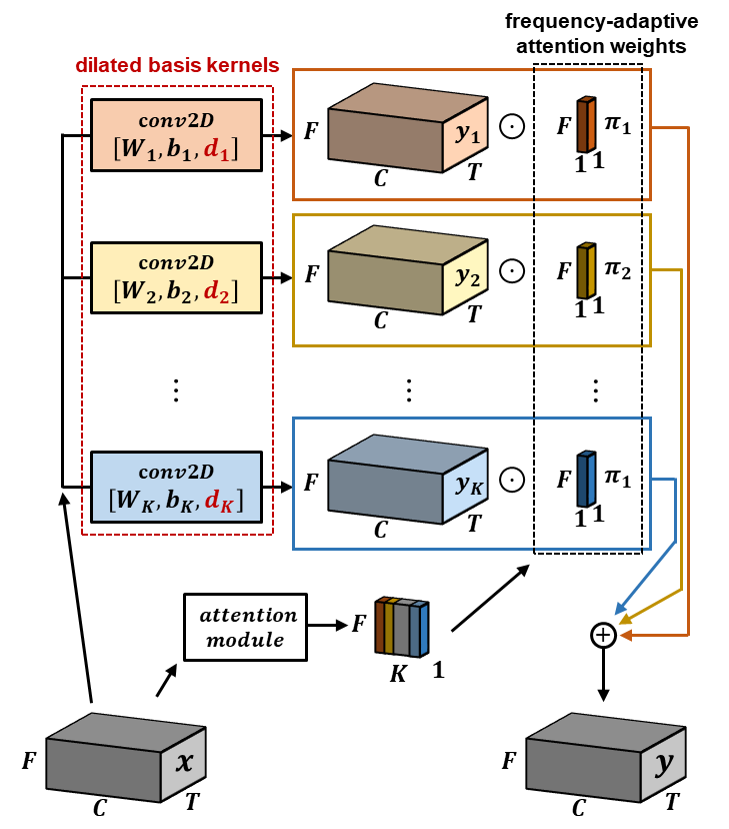}}
\vspace{-10pt}
\caption{An illustration of dilated frequency dynamic convolution operation. $x$ and $y$ are input and output of DFD conv. $K$ is number of basis kernels, $W_i$ and $b_i$, $d_i$ and $\pi_i$ are weight, bias, dilation size and frequency-adaptive attention weight for $i$-th basis kernel.}
\label{fig:dilatedFDYconv}
\vspace{-15pt}
\end{figure}

Frequency dynamic convolution (FDY conv) produces kernels varying over frequency by applying weighted sum on the basis kernels using frequency-adaptive attention weights \cite{FDY, dyconv}. It involves a presumption that basis kernels are trained to be dissimilar from each others through optimization process. However, if we assume an extreme case where basis kernels are trained to be the same, FDY conv will produce the same kernels over frequency no matter how different the frequency-adaptive attention weights are. Obviously this is a very unlikely case; the superior performance of FDY conv proves that FDY conv applies frequency-varying kernels. Nonetheless, there is no restriction on similarity between basis kernels within FDY conv mechanism, potentially limiting the performance. If basis kernels were forced to have different roles, we could improve SED by enhancing diversity of frequency-adaptive kernels. Meanwhile, there have been several attempts to expand the convolution receptive field for SED \cite{dilcrnn, dilesc, dwdilconv, dcase2021_1st}. Since sound events show time-frequency patterns those span broad spectro-temporal range, larger receptive field could improve SED. Especially, in CRNN structure, expanded receptive field along frequency dimension could help accurate detection of sound events since GRU is already used to take long-term temporal sound event patterns into account.

Therefore, to diversify frequency-adaptive kernel and expand its receptive field size at the same time, we propose \textit{dilated frequency dynamic convolution (DFD conv)} which applies varying dilation sizes on the basis kernels of FDY conv. Dilated convolution expands the receptive fields of convolution kernels without increasing number of trainable parameters by padding zeros within the kernels \cite{dilcrnn, dilesc, dilconv}. Therefore, by dilating the basis kernels of FDY conv, receptive fields of frequency-adaptive kernels will be expanded to recognize larger patterns without increasing the model size. In addition, applying different dilation sizes on basis kernels would train basis kernels to extract sound event information corresponding to different receptive field sizes. Thus role by each basis kernel would be specialized resulting in enhanced diversity of frequency-adaptive kernels. There has been similar approach using selective kernel (SK), which selects either 3 by 3 kernel or 5 by 5 kernel by attention. It has shown great SED performance by endowing diversity to kernel and introducing kernel with larger receptive field \cite{dcase2021_1st, sknet}. However, diversity of kernel is much smaller for SK which makes binary decision among two kernels while FDY conv and DFD conv makes new kernels by weighted-summing $K$ 3 by 3 basis kernels on each frequency bin. Furthermore, DFD conv could use expanded receptive field without increasing the model size with dilated basis kernels.

Detailed procedure of DFD conv is illustrated in Fig. \ref{fig:dilatedFDYconv}. $x$ and $y$ are input and output of DFD conv. $T$, $F$ and $C$ are input and output dimension size of time, frequency and channel. Note that FDY conv and DFD conv do not alter the input size. $W_i$, $b_i$, $d_i$ and $\pi_i$ are weight, bias, dilation size and frequency-adaptive attention weights for $i$-th basis kernel where $K$ is the number of basis kernels. Fig. \ref{fig:dilatedFDYconv} illustrates DFD conv operation where 2D convolution is performed on input with $K$ basis kernels to obtain $K$ intermediate outputs $y_i$ for $i=1,2, ... ,K$. Then final output $y$ is obtained by following equation:
\vspace{-5pt}
\begin{equation}
y_{tf}(x) = \sum_{i=1}^{K}\pi_{if}(x)y_{itf}
\vspace{-5pt}
\end{equation}
where $t$ is time and $f$ is frequency. While Fig. \ref{fig:dilatedFDYconv} shows different computation procedure compared to the figure in FDY conv paper \cite{FDY}, the actual algorithm follows equations on \cite{FDY} those are coherent to Fig. \ref{fig:dilatedFDYconv}.

\begin{table*}[ht]
\caption{Performance comparison between SED models using frequency dynamic convolution with various dilation sizes on four and five basis kernels on DESED real validation dataset. $d_{i,t}$ and $d_{i,f}$ implies dilation size of $i$th basis kernel in time and frequency dimensions respectively.}
\vspace{-10pt}
\centering
\setlength{\tabcolsep}{4.25pt}
\begin{tabular}{l|ll|ll|ll|ll|ll|l|ll}
\toprule
\textbf{models}        & $d_{1,t}$ & $d_{1,f}$ & $d_{2,t}$ & $d_{2,f}$ & $d_{3,t}$ & $d_{3,f}$ & $d_{4,t}$ & $d_{4,f}$ & $d_{5,t}$ & $d_{5,f}$ & \textbf{params (M)} & \textbf{PSDS1$\uparrow$} & \textbf{PSDS2$\uparrow$} \\
\midrule
FDY-CRNN (baseline)    & 1         & 1         & 1         & 1         & 1         & 1         & 1         & 1         & -         & -         & 11.061             & 0.441                    & 0.668                     \\
\midrule
DFD-CRNN, freq dilated & 1         & 1         & 1         & 1         & 1         & 1         & 1         & 2         & -         & -         & 11.061             & \textbf{0.444}           & \textbf{0.676}            \\
DFD-CRNN, time dilated & 1         & 1         & 1         & 1         & 1         & 1         & 2         & 1         & -         & -         & 11.061             & 0.442                    & 0.664                     \\
DFD-CRNN, both dilated & 1         & 1         & 1         & 1         & 1         & 2         & 2         & 1         & -         & -         & 11.061             & 0.442                    & 0.659                     \\
\midrule
DFD-CRNN, freq dilated & 1         & 1         & 1         & 1         & 1         & 1         & 1         & 1         & 1         & 2         & 12.628             & \textbf{0.445}           & \textbf{0.671}            \\
DFD-CRNN, time dilated & 1         & 1         & 1         & 1         & 1         & 1         & 1         & 1         & 2         & 1         & 12.628             & 0.442                    & 0.664                     \\
\bottomrule
\end{tabular}
\vspace{-10pt}
\label{tab:tf_results}
\end{table*}

\section{Experimental Setups}
\label{sec:setup}
\subsection{Implementation Details}
\vspace{-5pt}
Dataset used in this work is domestic environment sound event detection (DESED) composed of synthesized strongly labeled dataset, real weakly labeled dataset, real unlabeled dataset and real strongly labeled dataset \cite{DCASEtask4}. The former three are used for training and validation, while the last, named as real valiadation dataset, is used for evaluation as listed in Table \ref{tab:tf_results}, \ref{tab:size_results} and \ref{tab:improv_results}. DESED consists of 10-seconds-long audio files with sampling rate of 16 kHz. No external dataset is used in this work.

From the waveforms, we extract mel spectrogram with number of FFT of 2048, hop length of 256, Hamming window and 128 mel bins. Mel spectrograms are batched and fed into CRNN-based SED models. The baseline model in this work includes a CNN block with 7 convolution layers, where the first convolution layer uses conventional 2D convolution while the rest convolution layers use FDY conv, and is indicated as FDY-CRNN in the results. DFD convs only replace FDY convs from 2nd to 6th convolution layers for other models since input for the last layer has only 2 frequency bins and dilated kernel would be too large for that input. We indicated SED models with DFD conv as DFD-CRNN. Both FDY conv and DFD conv in SED models use $K=4$ as found optimal in \cite{FDY}. RNN block is composed of two bidirectional gated recurrent unit (GRU) layers. Then one fully connected (FC) layer is applied to obtain strong prediction while additional attention module is applied to obtain weak prediction \cite{DCASEtask4}.

We apply multiple data augmentation methods during training to maximize the effect by mean teacher method \cite{mytechreport, filtaug, meanteacher} as follows: frame shift \cite{dcasebaseline}, mixup \cite{mixup}, time masking \cite{specaug} and FilterAugment \cite{filtaug}. To exclude the effect of model-specific post-processing, we applied median filter with 7 frames (corresponding to 450ms) for all classes. We then calibrated median filtering on the best DFD-CRNN model. Note that SED model with FDY conv results different from the previous work as we adjusted seed, temperature and median filter.

\subsection{Evaluation Metrics}
\vspace{-5pt}
We use polyphonic sound detection score (PSDS) to evaluate SED performance \cite{PSDS}. PSDS decides correctness of predictions by considering the portion of intersection between the model prediction and the ground truth. In addition, it applies area under curve (AUC) - receiver operating characteristic (ROC) curves with multiple threshold points, minimizing the need of threshold optimization. While collar-based F1 score has been widely used, it decides if a prediction is correct based on the absolute time difference of onset and offset between the model prediction and ground truth \cite{sedmetrics}. This does not consider the length of sound events which might be too short or too long compared to the time difference criterion for the collar-based F1 score. In addition, F1 score requires tedious threshold optimization for better performance. Thus we use two types of PSDS used to rank detection and classification of acoustic scenes and events (DCASE) challenge 2021 \& 2022 task4 \cite{DCASEtask4}, where PSDS1 favors SED models with accurate time localization and PSDS2 favors SED models with accurate classification with less cross triggers. The performance comparison is based on PSDS1 + PSDS2 as DCASE challenges, and the PSDS results listed in Table \ref{tab:tf_results}, \ref{tab:size_results} and \ref{tab:improv_results} are the best score among 24 SED models from 12 separate training runs, including student models and teacher models by mean teacher method \cite{meanteacher}.

\subsection{Adapting Class-wise Median Filter}
\vspace{-5pt}
Since we perform experiments shown in Table \ref{tab:tf_results} and \ref{tab:size_results} with fixed median filter of length 7, we adapt class-wise median filter on the best DFD-CRNN for further improvement. We evaluate class-wise performance on the best DFD-CRNN using intersection-based F1 score with varying median filter sizes \cite{insightdcase2020}. While we used collar-based F1 score for adapting class-wise median filter in previous work \cite{FDY}, we used intersection-based F1 score in this work to make evaluation less dependent to onset and offset time differences. In addition, PSDS is based on intersection as well \cite{PSDS}. Then, we choose median filter size for each class that yields best intersection-based F1 scores for 24 SED models obtained by 12 separate training runs.

\section{Results and Discussion}
\label{sec:result}
\begin{table}[t]
\vspace{-5pt}
\caption{Performance comparison between SED models with larger kernels and dilated kernels along frequency dimension.}
\vspace{-10pt}
\centering
\setlength{\tabcolsep}{4.25pt}
\begin{tabular}{l|l|ll}
\toprule
\textbf{models}                  & \textbf{params (M)} & \textbf{PSDS1$\uparrow$} & \textbf{PSDS2$\uparrow$} \\
\midrule
CRNN                             & 4.428               & 0.410                    & 0.634                     \\
\midrule
+ kernel size = 3,5              & 5.472               & 0.423                    & 0.645                     \\
+ kernel size = 3,7              & 6.517               & 0.418                    & 0.649                     \\
\midrule
+ dilation size = 1,2            & 4.428               & 0.412                    & 0.643                     \\
+ dilation size = 1,3            & 4.428               & 0.404                    & 0.620                     \\
\bottomrule
\end{tabular}
\vspace{-15pt}
\label{tab:size_study}
\end{table}

\subsection{Preliminary experiments on receptive field sizes}
\vspace{-5pt}
To study the effect of receptive field size of convolution modules, we performed preliminary experiments on a CRNN model with conventional 2D convolution modules with larger kernel sizes and dilation sizes in frequency dimension. The results are shown in Table \ref{tab:size_study}, where larger kernels along frequency dimension improved the performance with larger receptive field. Dilated kernels with size 2 improved the performance while dilated kernel with size 3 rather worsened the performance. While larger kernels improved the performance, it involves huge increase in the model size. On the other hand, dilating kernels only marginally improved the performance while it does not involve more parameters. By applying DFD conv with varying dilation sizes on different basis kernels, we expect frequency-adaptive kernels which are weighted sum of basis kernels with different dilation sizes would have combination of the advantages of two different approaches; while the model size do not increase due to application of dilation, resultant frequency-adaptive kernels' receptive field will be as large as 3 by 5 or 3 by 7 kernels due to summation of basis kernels with different dilation sizes.

\begin{table*}[ht]
\caption{Performance comparison between SED models using frequency dynamic convolution with various dilation sizes on DESED real validation dataset. $d_{i,t}$ and $d_{i,f}$ implies dilation size of $i$th basis kernel in time and frequency dimensions respectively.}
\vspace{-10pt}
\centering
\setlength{\tabcolsep}{4.25pt}
\begin{tabular}{l|ll|ll|ll|ll|ll}
\toprule
\textbf{models}                 & \textbf{$d_{1,t}$} & \textbf{$d_{1,f}$} & \textbf{$d_{2,t}$} & \textbf{$d_{2,f}$} & \textbf{$d_{3,t}$} & \textbf{$d_{3,f}$} & \textbf{$d_{4,t}$} & \textbf{$d_{4,f}$} & \textbf{PSDS1$\uparrow$} & \textbf{PSDS2$\uparrow$} \\
\midrule
FDY-CRNN (baseline)             & 1                  & 1                  & 1                  & 1                  & 1                  & 1                  & 1                  & 1                  & 0.441                    & 0.668                     \\
\midrule
DFD-CRNN, dilation size = 2     & 1                  & 1                  & 1                  & 1                  & 1                  & 1                  & 1                  & 2                  & 0.444                    & \textbf{0.676}            \\
                                & 1                  & 1                  & 1                  & 1                  & 1                  & 2                  & 1                  & 2                  & \textbf{0.447}           & 0.675                     \\
                                & 1                  & 1                  & 1                  & 2                  & 1                  & 2                  & 1                  & 2                  & 0.442                    & 0.666                     \\
                                & 1                  & 2                  & 1                  & 2                  & 1                  & 2                  & 1                  & 2                  & 0.438                    & 0.665                     \\
DFD-CRNN, dilation size = 3     & 1                  & 1                  & 1                  & 1                  & 1                  & 1                  & 1                  & 3                  & 0.444                    & 0.673                     \\
                                & 1                  & 1                  & 1                  & 1                  & 1                  & 3                  & 1                  & 3                  & \textbf{0.447}           & 0.673                     \\
DFD-CRNN, dilation size = 4     & 1                  & 1                  & 1                  & 1                  & 1                  & 1                  & 1                  & 4                  & 0.443                    & 0.672                     \\
                                & 1                  & 1                  & 1                  & 1                  & 1                  & 4                  & 1                  & 4                  & 0.441                    & 0.674                     \\
\midrule
DFD-CRNN, varied dilation sizes & 1                  & 1                  & 1                  & 1                  & 1                  & 2                  & 1                  & 3                  & 0.442                    & 0.674                     \\
                                & 1                  & 1                  & 1                  & 2                  & 1                  & 2                  & 1                  & 3                  & \textbf{0.448}           & 0.672                     \\
DFD-CRNN (best)                 & 1                  & 1                  & 1                  & 2                  & 1                  & 3                  & 1                  & 3                  & \textbf{0.448}           & \textbf{0.688}            \\
                                & 1                  & 2                  & 1                  & 2                  & 1                  & 3                  & 1                  & 3                  & 0.447                    & 0.674                     \\
                                & 1                  & 1                  & 1                  & 2                  & 1                  & 3                  & 1                  & 4                  & 0.441                    & 0.672                     \\
\bottomrule
\end{tabular}
\vspace{-10pt}
\label{tab:size_results}
\end{table*}

\subsection{Experiments on Dilation Dimensions}
\vspace{-5pt}
We then experimented on the effect of dilation on frequency and time dimensions by comparing PSDS of DFD-CRNNs of which one basis kernel is dilated with size two along time or frequency dimension. Note that dilation size of one implies no dilation. In addition, we experimented DFD-CRNN with dilation on both time and frequency dimensions by applying one time-dilated basis kernel and one frequency-dilated basis kernel with dilation size two. The results are shown in Table \ref{tab:tf_results}, showing that while applying frequency-dilated basis kernel improves PSDS by 1\%, time-dilated and both-dilated basis kernels adversely affects the performance. Coherent to the previous presumption and the conclusion of section 4.1, dilation turns out to be only advantageous when applied on frequency dimension.

We further experimented with DFD-CRNN with five basis kernels to test if it is more beneficial to dilate new basis kernel rather than to dilate existing one. The results are illustrated in Table \ref{tab:tf_results} and those imply consistent results with the previous experiments; while adding frequency-dilated basis kernel improves performance, adding time-dilated basis kernel do not make significant difference. But the performance gain of adding dilated basis kernel over dilating existing basis kernel is not significant while it introduces additional parameters due to the added basis kernel. Therefore, we concluded that dilation improves SED performance when applied on frequency dimension rather than on time dimension, and it is more efficient to dilate existing basis kernel.

\begin{table}[t]
\vspace{-5pt}
\caption{Performance comparison between SED models with adapting class-wise median filters.}
\vspace{-10pt}
\centering
\setlength{\tabcolsep}{4.25pt}
\begin{tabular}{l|ll}
\toprule
\textbf{models}                 & \textbf{PSDS1$\uparrow$} & \textbf{PSDS2$\uparrow$} \\
\midrule
FDY-CRNN (baseline)             & 0.441                    & 0.668                     \\
FDY-CRNN + cw-mf \cite{FDY}      & 0.452                    & 0.670  
                         \\
\midrule
DFD-CRNN (best)                 & 0.448                    & 0.688                     \\
DFD-CRNN (best) + cw-mf          & \textbf{0.455}           & \textbf{0.702}            \\
\bottomrule
\end{tabular}
\vspace{-10pt}
\label{tab:improv_results}
\end{table}

\subsection{Experiments on Varying Dilation Sizes}
\vspace{-5pt}
Experiments on DFD-CRNN with various dilation sizes on frequency dimension are listed on Table \ref{tab:size_results}. Parameter counts are omitted as all models in the Table \ref{tab:size_results} share the same parameter counts. We first experimented on one, two, three and four basis kernels with dilation size of two in order to test the effect by multiple dilated basis kernels. The results shows that two basis kernels with dilation size of two on frequency dimension results the best. Applying dilation to all basis kernels decrease the performance, proving that applying varying dilation sizes to basis kernels improves SED performance by diversifying frequency-adaptive kernels. We tested on one or two basis kernels with dilation size of three and four. Results on Table \ref{tab:size_results} show that when applied to one or two basis kernels, dilation size of three is as good as dilation size of two but dilation size of four shows limited performance gain due to too sparse kernel.

Furthermore, we experimented applying varied dilation sizes on different basis kernels as shown in lower part of Table \ref{tab:size_results}. The best results is shown by one basis kernel with no dilation, one basis kernel with dilation size of two, and two basis kernels with dilation size of three, outperforming the baseline FDY-CRNN by 2.43\% in terms of PSDS. Although we expected that four basis kernel with all different dilation sizes from one to four would produce more diverse frequency-adaptive kernel, it turned out to be not as effective. This result agrees with the experiment of DFD-CRNN with dilation size of four which has shown insignificant improvement.

\begin{figure}[t]
\centerline{\includegraphics[width=8cm]{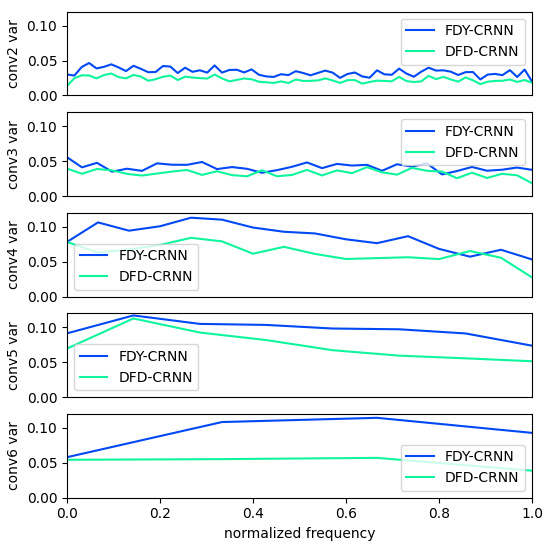}}
\vspace{-10pt}
\caption{Plots comparing variance of attention weights on 2nd - 6th convolution layers in FDY-CRNN and DFD-CRNN.}
\label{fig:attvar}
\vspace{-20pt}
\end{figure}

\subsection{Attention Weight Variance Comparison}
\vspace{-5pt}
To further analyze the effect of diversifying basis kernels with dilation, we compared variance of attention weight vectors in 2nd - 6th dynamic convolution layers of FDY-CRNN and DFD-CRNN with dilation size of 1, 2, 3 and 3 over frequency dimension as illustrated in Fig. \ref{fig:attvar}. 
To obtain the variance, foreground train soundbank of DESED were inputted to FDY-CRNN and DFD-CRNN, then distance variance of obtained attention weight vectors $\mathbf{w} \in \mathbb{R}^K$, which is composed of attention weights from the attention module in Fig. \ref{fig:dilatedFDYconv} along single frequency bin, were calculated on each frequency bin of each convolution layer using following equation:
\vspace{-5pt}
\begin{equation}
\mathrm{var}_{lf} = \frac{1}{N}\sum_{i=1}^{N} \left\Vert \frac{1}{N}\sum_{j=1}^{N} \mathbf{w_{jlf}} - \mathbf{w_{ilf}} \right\Vert _2 ^2
\vspace{-5pt}
\end{equation}
\noindent where $l$ is convolution layer index, $f$ is frequency index and $N$ is total number of data in foreground train soundbank. Except few exceptions on 3rd and 4th convolution layers, DFD-CRNN consistently shows smaller variance than FDY-CRNN does. Since DFD-CRNN outperforms FDY-CRNN, we can infer that basis kernels of DFD conv effectively perform diverse roles thus less variance in attention weights was sufficient for frequency-adaptive kernel of DFD conv to perform SED with better quality. Thus we conclude that basis kernels in DFD conv are effectively diversified thus DFD conv performs SED more effectively with less change in attention weights.

\subsection{Class-wise Median Filter}
\vspace{-5pt}
Adapting class-wise median filter to best DFD-CRNN model resulted in model outperforming previous work \cite{FDY} by 3.12\%  as shown in Table \ref{tab:improv_results}. While class-wise median filter adaptation improves PSDS1 better with collar-based F1 score as in FDY-CRNN \cite{FDY}, it improves PSDS2 better with intersection-based F1 score. It is because collar-based F1 score focuses more on time-localization of onset and offset when evaluating SED.

\section{Conclusion}
\label{sec:conclusion}
We proposed \textit{dilated frequency dynamic convolution (DFD conv)} to expand and diversify FDY conv for SED. Experimental results revealed that dilation is more effective in frequency dimension than in time dimension on CRNN structure. In addition, varying dilation sizes for different basis kernels further improved performance. Analysis on attention weight vector variance has shown that DFD conv uses more diverse convolution kernels than FDY conv does. Best DFD-CRNN outperforms the baseline FDY-CRNN by 3.12\% with adapted class-wise median filter in terms of PSDS.

\bibliographystyle{IEEEtran}
\bibliography{mybib}

\begin{thebibliography}{10}
\providecommand{\url}[1]{#1}
\csname url@samestyle\endcsname
\providecommand{\newblock}{\relax}
\providecommand{\bibinfo}[2]{#2}
\providecommand{\BIBentrySTDinterwordspacing}{\spaceskip=0pt\relax}
\providecommand{\BIBentryALTinterwordstretchfactor}{4}
\providecommand{\BIBentryALTinterwordspacing}{\spaceskip=\fontdimen2\font plus
\BIBentryALTinterwordstretchfactor\fontdimen3\font minus \fontdimen4\font\relax}
\providecommand{\BIBforeignlanguage}[2]{{%
\expandafter\ifx\csname l@#1\endcsname\relax
\typeout{** WARNING: IEEEtran.bst: No hyphenation pattern has been}%
\typeout{** loaded for the language `#1'. Using the pattern for}%
\typeout{** the default language instead.}%
\else
\language=\csname l@#1\endcsname
\fi
#2}}
\providecommand{\BIBdecl}{\relax}
\BIBdecl

\bibitem{CASSE}
T.~Virtanen, M.~D. Plumbley, and D.~Ellis, \emph{Computational Analysis of Sound Scenes and Events}, 1st~ed.\hskip 1em plus 0.5em minus 0.4em\relax Springer Publishing Company, Incorporated, 2017, pp. 3--11, 71--77.

\bibitem{DCASEtask4}
N.~Turpault, R.~Serizel, A.~Parag~Shah, and J.~Salamon, ``{Sound event detection in domestic environments with weakly labeled data and soundscape synthesis},'' in \emph{{Workshop on Detection and Classification of Acoustic Scenes and Events}}, 2019.

\bibitem{crnn}
E.~{\c{C}}ak{\i}r, G.~Parascandolo, T.~Heittola, H.~Huttunen, and T.~Virtanen, ``Convolutional recurrent neural networks for polyphonic sound event detection,'' \emph{IEEE/ACM Transactions on Audio, Speech, and Language Processing}, vol.~25, no.~6, pp. 1291--1303, 2017.

\bibitem{sedmetrics}
A.~Mesaros, T.~Heittola, and T.~Virtanen, ``Metrics for polyphonic sound event detection,'' \emph{Applied Sciences}, vol.~6, no.~6, 2016.

\bibitem{mytechreport}
H.~Nam, B.-Y. Ko, G.-T. Lee, S.-H. Kim, W.-H. Jung, S.-M. Choi, and Y.-H. Park, ``Heavily augmented sound event detection utilizing weak predictions,'' DCASE2021 Challenge, Tech. Rep., 2021.

\bibitem{FDY}
H.~Nam, S.-H. Kim, B.-Y. Ko, and Y.-H. Park, ``{Frequency Dynamic Convolution: Frequency-Adaptive Pattern Recognition for Sound Event Detection},'' in \emph{Proc. Interspeech}, 2022.

\bibitem{filtaug}
H.~Nam, S.-H. Kim, and Y.-H. Park, ``Filteraugment: An acoustic environmental data augmentation method,'' in \emph{International Conference on Acoustics, Speech and Signal Processing (ICASSP)}, 2022.

\bibitem{freqatt}
H.~Nam, S.-H. Kim, D.~Min, and Y.-H. Park, ``Frequency \& channel attention for computationally efficient sound event detection,'' in \emph{{Workshop on Detection and Classification of Acoustic Scenes and Events}}, 2023.

\bibitem{dcase2023_1st}
J.~W. Kim, S.~W. Son, Y.~Song, H.~K. Kim, I.~H. Song, and J.~E. Lim, ``Semi-supervised learning-based sound event detection using frequency dynamic convolution with large kernel attention for {DCASE} challenge 2023 task 4,'' DCASE2023 Challenge, Tech. Rep., 2023.

\bibitem{dcase2023_2nd}
S.~Xiao, J.~Shen, A.~Hu, X.~Zhang, P.~Zhang, and Y.~Yan, ``Sound event detection with weak prediction for dcase 2023 challenge task4a,'' DCASE2023 Challenge, Tech. Rep., 2023.

\bibitem{PSDS}
{\c{C}}.~Bilen, G.~Ferroni, F.~Tuveri, J.~Azcarreta, and S.~Krstulovi\'{c}, ``A framework for the robust evaluation of sound event detection,'' in \emph{International Conference on Acoustics, Speech and Signal Processing (ICASSP)}, 2020, pp. 61--65.

\bibitem{dyconv}
Y.~Chen, X.~Dai, M.~Liu, D.~Chen, L.~Yuan, and Z.~Liu, ``Dynamic convolution: Attention over convolution kernels,'' in \emph{IEEE/CVF Conference on Computer Vision and Pattern Recognition (CVPR)}, June 2020.

\bibitem{dilcrnn}
Y.~Li, M.~Liu, K.~Drossos, and T.~Virtanen, ``Sound event detection via dilated convolutional recurrent neural networks,'' in \emph{International Conference on Acoustics, Speech and Signal Processing (ICASSP)}, 2020.

\bibitem{dilesc}
X.~Zhang, Y.~Zou, and W.~Shi, ``Dilated convolution neural network with leakyrelu for environmental sound classification,'' in \emph{International Conference on Digital Signal Processing}, 2017.

\bibitem{dwdilconv}
K.~Drossos, S.~I. Mimilakis, S.~Gharib, Y.~Li, and T.~Virtanen, ``Sound event detection with depthwise separable and dilated convolutions,'' in \emph{International Joint Conference on Neural Networks}, 2020.

\bibitem{dcase2021_1st}
X.~Zheng, H.~Chen, and Y.~Song, ``Zheng ustc team’s submission for dcase2021 task4 – semi-supervised sound event detection,'' DCASE2021 Challenge, Tech. Rep., 2021.

\bibitem{dilconv}
F.~Yu and V.~Koltun, ``Multi-scale context aggregation by dilated convolutions,'' in \emph{International Conference on Learning Representations (ICLR)}, 2016.

\bibitem{sknet}
X.~Li, W.~Wang, X.~Hu, and J.~Yang, ``Selective kernel networks,'' in \emph{IEEE/CVF Conference on Computer Vision and Pattern Recognition (CVPR)}, 2019.

\bibitem{meanteacher}
A.~Tarvainen and H.~Valpola, ``Mean teachers are better role models: Weight-averaged consistency targets improve semi-supervised deep learning results,'' in \emph{Advances in Neural Information Processing Systems}, vol.~30, 2017.

\bibitem{dcasebaseline}
\BIBentryALTinterwordspacing
N.~Turpault. Dcase2021 task4 baseline. GitHub. Available: https://github.com/DCASE-REPO/DESED\_task. [Online]. Available: \url{https://github.com/DCASE-REPO/DESED\_task}
\BIBentrySTDinterwordspacing

\bibitem{mixup}
H.~Zhang, M.~Cisse, Y.~N. Dauphin, and D.~Lopez-Paz, ``mixup: Beyond empirical risk minimization,'' in \emph{International Conference on Learning Representations}, 2018.

\bibitem{specaug}
D.~S. Park, W.~Chan, Y.~Zhang, C.-C. Chiu, B.~Zoph, E.~D. Cubuk, and Q.~V. Le, ``{SpecAugment: A Simple Data Augmentation Method for Automatic Speech Recognition},'' in \emph{Proc. Interspeech}, 2019.

\bibitem{insightdcase2020}
G.~Ferroni, N.~Turpault, J.~Azcarreta, F.~Tuveri, R.~Serizel, {\c{C}a\v{g}da\c{s}}.~Bilen, and S.~Krstulovi\'{c}, ``Improving sound event detection metrics: Insights from dcase 2020,'' in \emph{International Conference on Acoustics, Speech and Signal Processing (ICASSP)}, 2021.

\end{thebibliography}

\end{document}